\documentclass[amsfonts]{article}
\usepackage{graphicx}
\pdfoutput=1
\usepackage{subcaption}

\newcommand{\be}{\begin{equation}}
\newcommand{\ee}{\end{equation}}
\newcommand{\bea}{\begin{array}}
\newcommand{\ea}{\end{array}}
\newcommand{\beqa}{\begin{eqnarray}}
\newcommand{\eeqa}{\end{eqnarray}}

\newcommand{\eean}{\end{eqnarray*}}

\def\up#1{\leavevmode \raise.16ex\hbox{#1}}

\newcommand{\gapproxeq}{\lower
 .7ex\hbox{$\;\stackrel{\textstyle >}{\sim}\;$}}
\newcommand{\lapproxeq}{\lower .7ex\hbox{$\;\stackrel
{\textstyle <}{\sim}\;$}}

\newcounter{appendice}

\def\thebibliography#1{{\bf REFERENCES\markboth
 {REFERENCES}{REFERENCES}}\list
 {[\arabic{enumi}]}{\settowidth\labelwidth{[#1]}\leftmargin\labelwidth
 \advance\leftmargin\labelsep
 \usecounter{enumi}}
 \def\newblock{\hskip .11em plus .33em minus -.07em}
 \sloppy
 \sfcode`\.=1000\relax}

\def\BI{{\rm 1\!l}}
\begin{document}
\centerline{ \LARGE Lorentzian Fuzzy Spheres}

\vskip 2cm

\centerline{A. Chaney\footnote{adchaney@crimson.ua.edu},    	Lei Lu\footnote{llv1@crimson.ua.edu} and A. Stern\footnote{astern@ua.edu}   }

\vskip 1cm
\begin{center}
  { Department of Physics, University of Alabama,\\ Tuscaloosa,
Alabama 35487, USA\\}

\end{center}
\vskip 2cm

\vspace*{5mm} 

\normalsize
\centerline{\bf ABSTRACT}

We show that fuzzy spheres are solutions of {\it Lorentzian} IKKT  matrix models.    The  solutions serve as  toy models of  closed noncommutative cosmologies where  big bang/crunch singularities appear only after taking the commutative limit.  The commutative limit of these solutions corresponds to a sphere embedded in Minkowski space.  This `sphere'  has several novel features.  The induced metric does not agree with the standard metric on the sphere, and moreover, it does not have a fixed signature.  The curvature computed from the induced metric is not  constant, has singularities at fixed latitudes (not corresponding to the poles)  and  is  negative. Perturbations  are made about the solutions, and are shown to yield  a scalar field theory on the sphere in the commutative limit. The scalar field can become tachyonic for a range of the  parameters of the theory.

\bigskip
\bigskip

\newpage

\section{Introduction}

It is well known that fuzzy spheres and fuzzy coset spaces\cite{Madore:1991bw}-\cite{Iso:2001mg}  are solutions to  matrix models.  More specifically, they are solutions to the bosonic sector of  {\it  Euclidean}  Ishibashi, Kawai, Kitazawa,  Tsuchiya (IKKT) matrix models.\cite{Ishibashi:1996xs} 
  An application of these solutions to particle physics has been to make extra dimensions noncommutative.\cite{Aschieri:2003vy}  Here we show that fuzzy spheres can also be solutions to IKKT matrix models with a Minkowski background metric tensor.  This means  that  in addition to making  extra dimensions noncommutative, fuzzy spheres and more generally fuzzy coset spaces can  be used to make space-time noncommutative.  Moreover, they can be  toy models for noncommutative cosmological space-times.

Various aspects of  Lorentzian IKKT matrix models have been discussed in the literature, including classical solutions  and their implications for cosmology.\cite{Klammer:2009ku},\cite{Steinacker:2011ix},\cite{Kim:2011cr},\cite{Jurman:2013ota},\cite{Stern:2014uea} The solutions were generally written in terms of infinite-dimensional matrices, and they may or may not be associated with finite dimensional Lie algebras.   The advantage of now having fuzzy spheres and  fuzzy coset spaces as solutions is that they can be expressed in terms of $N\times N$ matrices, where  $N$ is finite, and they lead to closed space-time cosmologies upon taking the $N\rightarrow\infty$ limit, corresponding to the commutative limit. Big bang/crunch singularities should then appear in this limit, while the finite dimensional matrix  description is singularity free. 

In section two we write down a fuzzy sphere solution to the Lorentzian IKKT model.  The model is written down specifically  in three space-time dimensions and cubic and quadratic terms are included in the action. We show that the solution yields  a closed (two-dimensional) universe in the commutative limit.  While the commutative limit of the solution  is topologically a two-sphere, there are  a number of novel features, arising from the fact that it is embedded in a three-dimensional Minkowski space. 
  The induced metric  does not agree with the standard metric on the sphere, and moreover, it does not have a fixed signature.  The curvature computed from the induced metric is not  constant and it is  negative.  It is singular at two fixed latitudes (which are not located at the poles) and time-like geodesics originate and terminate at these latitudes.  Thus in this toy model, the big bang/crunch singularities occur at nonzero spatial size.

We examine perturbations around the fuzzy sphere solution in section three.  In the commutative limit, the perturbations are described by a scalar field coupled to a gauge field.  The latter can be eliminated yielding a scalar field which can propagate on the Lorentzian region of the two-dimensional surface.  Depending on the choice of parameters the scalar field can be massive, massless or tachyonic.

Concluding remarks are given in section four.
\section{Fuzzy sphere solution to the Lorentzian IKKT matrix model}

The setting  here is the bosonic sector of the Lorentzian IKKT matrix model  in three space-time dimensions.  The dynamical degrees of freedom for the matrix model are contained in three infinite-dimensional Hermitean matrices $X^\mu$, $\mu=0,1,2$, with $\mu=0$ indicating a time-like direction.  In addition to the standard Yang-Mills term, we  include a  cubic term and a quadratic term in the action (which are both necessary for obtaining fuzzy sphere solutions)
\beqa S(X)&=&\frac 1{g^2}{\rm Tr}\Bigl(-\frac 14 [X_\mu, X_\nu] [X^\mu,X^\nu] +\frac 23 i \alpha \epsilon_{\mu\nu\lambda} X^\mu X^\nu X^\lambda+\frac \beta 2X_\mu X^\mu\Bigr)\;,\label{mmactnplsqd}\eeqa
where $g$,  $\alpha$ and $\beta$ are real  coefficients.  Our conventions are $\epsilon_{012}=1$, and we raise and lower indices $\mu,\nu,...$ with the flat metric $[\eta_{\mu\nu}]=$diag$(-1,1,1)$.  
The  resulting equations of motion  are
\be [ [X_\mu,X_\nu],X^\nu]+i\alpha \epsilon_{\mu\nu\lambda}[X^\nu,X^\lambda] =-\beta X_\mu\label{eomwthqdtrm}\ee
The dynamics is invariant under three-dimensional
 Lorentz transformations, $X^\mu\rightarrow L^\mu_{\;\;\nu} X^\nu$,  where $L$ is a $3\times 3$ Lorentz matrix, and
unitary `gauge' transformations, $X^\mu\rightarrow UX^\mu U^\dagger$, where $U$ is an infinite dimensional unitary matrix. 
The equations of motion also have discrete symmetries, namely proper reflections.  An example is
\be (X^0,X^1,X^2)\rightarrow  (-X^0,X^1,-X^2)  \label{dscrtsmtre}\ee
 Translation invariance in the three-dimensional Minkowski space is broken when $\beta\ne 0$.

When $\beta \ne 0$, there exist finite dimensional matrix solutions to the equations of motion (\ref{eomwthqdtrm}), which are associated with the $su(2)$ algebra in an $N-$dimensional representation. Say the latter is spanned by $N\times N$ hermitean matrices $J_i,\;i=1,2,3$, satisfying  $[J_i,J_j]=i\alpha\epsilon_{ijk} J_k$.\footnote{The Levi-Civita symbol here is associated with Euclidian space, unlike the ones appearing in (\ref{mmactnplsqd}) and (\ref{eomwthqdtrm}) which are associated with Minkowski space.}
Let us set
  \be 
X^0=\frac{w_3}{\alpha} J_3\qquad 
X^1=\frac{w_1}{\alpha} J_1\qquad 
X^2=\frac{w_2}{\alpha}  J_2\;,\qquad\;\;\label{gnrlanstz}\ee
where $w_i$ are real.
Upon substituting this expression into the equations of motion one gets
\beqa (w_1^2+ w_2^2+\beta) w_3+2\alpha w_1w_2 &=&0 \cr & &  \cr
(w_2^2- w_3^2+\beta) w_1-2\alpha w_2w_3 &=&0  \cr & &  \cr
(w_1^2- w_3^2+\beta) w_2-2\alpha w_1w_3 &=&0 \;,
\eeqa
which has nontrivial solutions. Lorentz symmetry is in general broken by the solutions, unlike the case with de Sitter and anti-de Sitter solutions.\cite{Jurman:2013ota},\cite{Stern:2014aqa}   The $su(2)$ Casimir operator for any of the solutions can be written as
$ \frac 1{w_3^2}(X^0)^2+  \frac 1{w_1^2}(X^1)^2+ \frac 1{w_2^2}(X^2)^2$, which has the value $\frac 1 4({N^2-1})$ in the $N-$dimensional representation, thereby defining a fuzzy sphere,  or actually fuzzy ellipsoid since rotational invariance in the $(X^0,X^1,X^2)$ space  does not in general hold.

We note that the solution is invariant under the three-dimensional rotation group (rather than the Lorentz group) in the special case where   $w_1^2=w_2^2=w_3^2$.  Let us more generally restrict to the case of   rotational  invariance in the $(X^1,X^2)$ plane, which  means  $w_1^2=w_2^2$.  Two simple solutions exist in this case: 
  \be 
X^0=2  J_3\qquad 
X^1=\frac{\sqrt{-\beta}}{\alpha}\, J_1\qquad 
X^2=-\frac{\sqrt{-\beta}}{\alpha}\, J_2\;,\qquad\;\;\label{rtnvfntsln}\ee
\centerline {and}
 \be 
X^0=-2  J_3\qquad 
X^1=\frac{\sqrt{-\beta}}{\alpha}\, J_1\qquad 
X^2=\frac{\sqrt{-\beta}}{\alpha}\, J_2\;,\qquad\;\;\label{rtnvfntsln2}\ee
Nontrivial solutions require the presence of both the cubic and quadratic terms in (\ref{mmactnplsqd}),  $\alpha\ne 0$ and $\beta<0$.  (\ref{rtnvfntsln}) and (\ref{rtnvfntsln2})   are equivalent due to the discrete symmetry  (\ref{dscrtsmtre}).  For the sake of definitess we choose to work with the former (\ref{rtnvfntsln}).
The $su(2)$ Casimir operator for this solution can be written as
\be -\frac{\beta}{4\alpha^2} (X^0)^2+ (X^1)^2+(X^2)^2\label{su2casimir} \;,\ee
having the value $-\frac\beta 4({N^2-1})$ in the $N-$dimensional representation.
The `time' matrix $X^0$ then has discrete eigenvalues $ 2\alpha m$,  where $ m =\frac{ -N+1}2,\frac{-N+3}2,...,\frac{ N-1}2$.  For any $m$ defining a time-slice we can also define a spatial size.  Call $A$ the `space' matrix, where $A^2=X_+X_-$ and $X_\pm=X^1\pm iX^2$.  We can  identify it with  $-\frac{\beta}{\alpha^2} (\vec J^2-J_3^2-J_3)$ for the solution  (\ref{rtnvfntsln}).   $A^2$ then commutes with $X^0$ and has
eigenvalues
$  \; -\beta\Bigl(\frac {N^2-1}4-m^2-m\Bigr)$.  Thus time and the spatial size are discrete.
Examples of  spectra for $(X^0,A^2)$ for some $N-$dimensional  representations are\beqa
N=2&\quad &(-\alpha,-\beta),\;  (\alpha,0)\cr 
N=3&\quad &(-2\alpha,-2\beta) ,\;(\alpha,-\beta),\; (2\alpha,0)\cr 
N=4&\quad & (-3\alpha,-3\beta),\;(-\alpha,-4\beta),\; (\alpha,-3\beta),\;(3\alpha,0)\cr  
N=5&\quad &(-4\alpha,-4\beta),\;(-2\alpha,-6\beta),\;(\alpha,-6\beta),\; (2\alpha,-4\beta),\; (4\alpha,0) \eeqa
Say $\alpha>0$.  Then for large $N$, the  spatial size operator $A$ has eigenvalue $\sqrt{-\beta N}$ for   the lowest time  eigenvalue $\sim -\alpha N$, i.e., the initial state. It then increases to the maximum value $\sqrt{-\beta }\,N/2$ as the time goes to  zero, and then decreases to zero upon approaching at highest time eigenvalue $\sim\alpha N$, i.e.,  the final state.  This solution can thus be regarded as a discrete analogue of a closed  cosmological space-time.  The eigenvalues of $X^0$ versus those of $A$ are plotted for $N=100$  in figure 1. 
\begin{figure}[placement h]
\begin{center}
\includegraphics[height=2.5in,width=2.25in,angle=0]{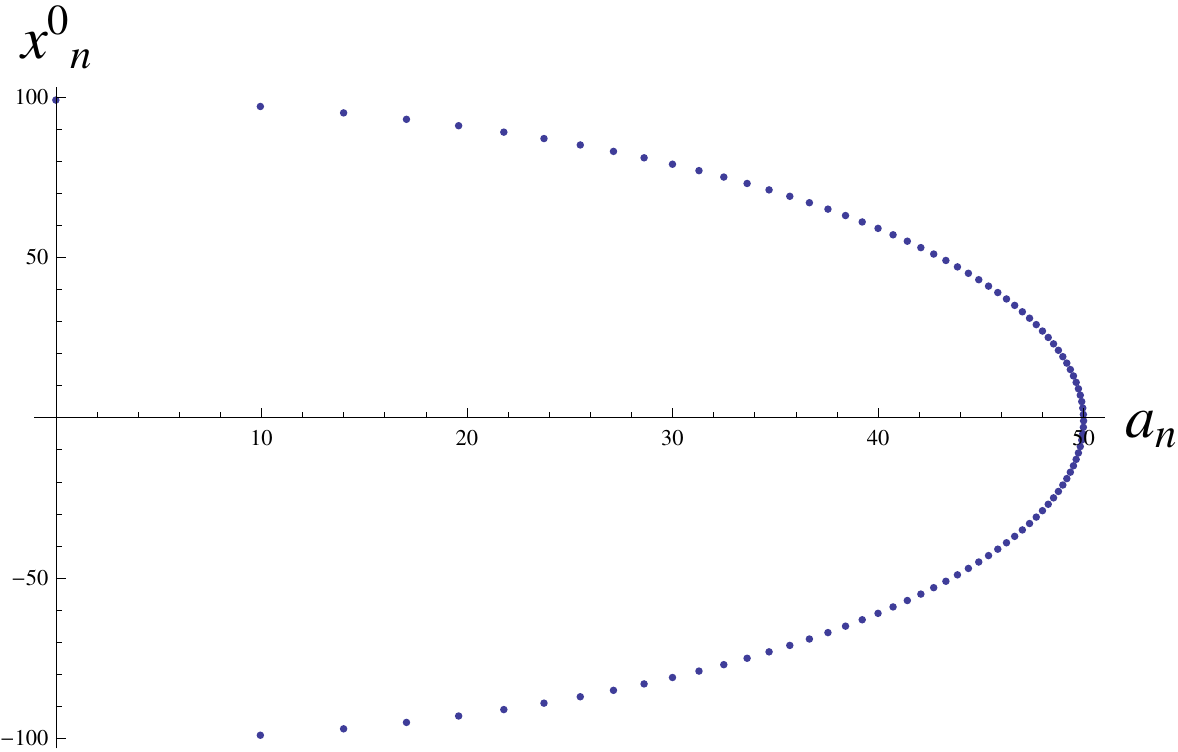}
\caption {Fuzzy closed universe solution.  Plot of  the eigenvalues $x^0_n$ of  the time matrix $X^0$ versus the eigenvalues $a_n$ of the space matrix  $A$ for  $N=100$, $\alpha= 1$ and $\beta=-1$.
  } 
\end{center}
\end{figure}

  Just as with the fuzzy sphere in a Euclidean background, the commutative limit of the matrix solution here is obtained by taking $N\rightarrow \infty$.  Here we also need $\alpha,\beta\rightarrow 0$, with $\alpha N$ and $\sqrt{-\beta }\,N$  finite in the limit.  The commutative limit of the solution is then characterized by two real parameters, which we denote by $a_0$ and $r^2$,
\be \frac{\sqrt{-\beta}}{2\alpha}\rightarrow a_0\qquad\qquad  \frac{\sqrt{-\beta}N}{2}\rightarrow r\label{eighttwo}\ee
One standardly defines the commutative limit in analogous fashion to the classical limit of a quantum theory, where we can take $\alpha$ to play the analogous role to $\hbar$. This means replacing the matrices $X^\mu$ by  space-time commuting coordinates which we denote by $x^\mu$,  where $x^0$ and $x^i,\;i=1,2$ denote the time and space coordinates, respectively.    The constraint on the $su(2)$  Casimir operator  (\ref{su2casimir}) means that in the commutative limit the solution satisfies.
\be a_0^2  (x^0)^2+ (x^1)^2+(x^2)^2=r^2 \;\label{sprcnstrnt}\ee
While real $a_0$ means that the solution is topologically a two sphere, there  are a number of novel features, which we show below, due to the fact that this `sphere' is embedded in Minkowksi space-time.

The commutative limit also requires replacing  the commutator of  functions of  $X^\mu$,  evaluated for the solution (\ref{rtnvfntsln}),  by $i\alpha$ times Poisson bracket of the same functions of the coordinates $x^\mu$.   Thus the Poisson brackets of the embedding coordinates are
\be \{x^0,x^1\}=-2 x^2\qquad\quad  \{x^2,x^0\}=-2 x^1\qquad\quad  \{x^1,x^2\}=-2a_0^2 x^0\ee
We can express $x^\mu$ in terms of angular momenta $j_i,\;i=1,2,3$, which satisfies the $su(2)$   Poisson bracket algebra  $\{j_i  ,j_j\}=\epsilon_{ijk} j_k$, using 
\be \pmatrix{x^0\cr x^1\cr x^2}=2 \pmatrix{j_3 \cr a_0 j_1 \cr - a_0  j_2}\;,\label{s3ncprtztn}\ee and from (\ref{sprcnstrnt}), $j_1^2+j_2^2+j_3^2=(\frac r{2a_0})^2$. For simplicity, we set $r=2 a_0$ so that $j_i$ spans a sphere of unit radius.   We can introduce standard spherical coordinates  $ (\theta,\phi)$, $0< \theta< \pi,\;0\le\phi<2\pi$,  and write
\be j_1=\sin\theta\cos\phi\qquad \quad   j_2=\sin\theta\sin\phi\qquad \quad   j_3=\cos\theta \;\ee
 The $su(2)$   Poisson bracket algebra for $j_i$ is recovered upon defining the Poisson brackets on the sphere to be
\be  \{ F,{ G} \}(\theta, \phi)=\csc\theta \,\Bigl( \partial_\theta{ F} \partial_\phi { G}-\partial_\phi { F} \partial_\theta { G} \Bigr)\label{sixty8}\;,\ee for any two functions $F$ and $G$ on the sphere.  

The induced metric ${\tt g}_{ab}=\partial_a x^\mu\partial_b x_\mu,$ $a,b,...=\theta,\phi$, does not agree with the standard metric on the sphere, and moreover, it does not have a fixed signature.  The curvature computed from the induced metric is not  constant and  is  negative!
The invariant interval constructed from the induced metric is
\be -d\tau^2=4\Bigl(a_0^2 \cos^2\theta-\sin^2\theta \Bigr)\,d\theta^2 +4a_0^2\sin^2\theta\, d\phi^2  \label{metricons2}\ee
${\tt g}_{\theta\theta}$ vanishes at two latitudes $\theta=\theta_\pm$ on the sphere defined by $\tan\theta_\pm=|a_0|$.  Say that $\theta=\theta_+$ is contained  in the  northern hemisphere,  $0<\theta_+<\frac\pi2$, while $\theta=\theta_-$ is contained  in  the southern hemisphere, $\frac\pi2<\theta<\pi$.   The signature on sphere  is Euclidean for  $0<\theta<\theta_+$ and $\theta_-<\theta<\frac \pi 2$, while it is Lorentzian for  $\theta_+<\theta <\theta_-$.   We can regard $\theta$ as a time-like variable for the latter, with  $2 a_0\sin\theta$ being the spatial radius at any time-slice.   $\theta=\theta_\pm$ correspond to singularities in the curvature, and are not coordinate singularities. The Ricci scalar computed from the induced metric is
\be R=-\frac {1}{2(a_0^2\cos\theta^2-\sin^2\theta)^2}\;,\label{crvtr}\ee
and thus it is singular at  the latitudes $\theta=\theta_\pm$.   (\ref{crvtr}) shows that  the curvature in the nonsingular regions is everywhere negative.  The singularities of the Ricci tensor are analogous to big bang/crunch singularities, with the distinction that they occur at a nonzero spatial radius $2 a_0\sin\theta_\pm=\frac{2 a_0^2}{\sqrt{a_0^2+1}}$ on the two-dimensional space-time.
 Time-like longitudinal geodesics  originate and terminate at the singular latitudes $\theta=\theta_\pm$.  This is because their tangent vectors 
  $\;(\frac {d\theta}{d\tau},\frac {d\phi}{d\tau})=\Bigl(\frac {1}{\sqrt{\sin^2\theta - a_0^2\cos\theta^2}}\,,\,0\Bigr)\;$
are well defined  in the Lorentzian region, $\theta_+<\theta <\theta_-$, while they are imaginary in the  Euclidean regions,  $0<\theta<\theta_+$ and $\theta_-<\theta<\frac \pi 2$.  The total elapsed proper time along these geodesics is finite and given by the elliptic integral $\;2\int_{\tan^{-1}{a_0}}^{\pi -\tan^{-1}{a_0}}d\theta{\sqrt{\sin^2\theta - a_0^2\cos\theta^2}}\;$.

\section{Emergent  field dynamics}

Here we perturb around the matrix solution (\ref{rtnvfntsln}).    Similar to \cite{Iso:2001mg},  we can define  noncommutative field strengths $F_{\mu\nu}$ on the fuzzy sphere.  Here we take 
\beqa  
 F^{01}&=&\frac 1{\alpha}[X^0,X^1] +2i X^2\cr &&\cr
 F^{02}&=&\frac 1{\alpha}[X^0,X^2] -2i X^1\cr &&\cr
 F^{12}&=&\frac 1{\alpha}[X^1,X^2] -\frac{i\beta}{2\alpha} X^0\label{mfsfzysph}
 \;, \eeqa which transform covariantly under
unitary gauge transformations, $F_{\mu\nu}\rightarrow UF_{\mu\nu} U^\dagger$, and vanish when evaluated on the fuzzy sphere solutions (\ref{rtnvfntsln}).  The matrix action
 (\ref{mmactnplsqd}) can then be re-expressed in terms of the noncommutative field strengths
\beqa {g^2}S(X)&=&{\rm Tr}\biggl\{-\frac{\alpha^2}4 F_{\mu\nu}F^{\mu\nu} -\frac 43 i \alpha^2(F^{01}X^2+F^{20}X^1) + i \alpha^2\Bigl(  \frac 23 -\frac\beta{2\alpha^2}\Bigr)F^{12}X^0   \cr &&\cr &&\qquad+\Bigl(\frac \beta 2-\frac{2\alpha^2}3\Bigr)\Bigl((X^1)^2+(X^2)^2\Bigr)+\beta\Bigl(\frac \beta{8\alpha^2} -\frac 56\Bigr)(X^0)^2\biggr\}\label{ninefive}\eeqa
Now perturb around the matrix solution (\ref{rtnvfntsln}) using
\be X^0=2  \Bigl( J_3+ \frac {\alpha^2}{\sqrt{-\beta}} A^0 \Bigr) \qquad\quad  X^1=\frac{\sqrt{-\beta}}\alpha\, J_1+ \alpha A^1 \qquad\quad  X^2=-\frac{\sqrt{-\beta}}\alpha\, J_2-\alpha A^2\label{one0three}\;, \ee
where the perturbations are functions on the fuzzy sphere,  $A^\mu= A^\mu(J_1,J_2,J_3)$.
If we write infinitesimal unitary gauge transformations using $U=\BI-\frac{i\alpha}{\sqrt{-\beta}}\Lambda $, where $\Lambda$ is a hermitean matrix with infinitesimal elements, then the infinitesimal variations of $A^\mu$ read
\beqa  \delta A^\mu &=&- i\Bigl(\frac 1\alpha [\Lambda,   J^\mu] +\frac{ \alpha}{\sqrt{-\beta}}[\Lambda ,A^\mu]\Bigr) \;,\label{1Ofour} \eeqa where we identify $(J^0,J^1,J^2)$ with  $(J_3,J_1,J_2)$.
Substituting  (\ref{one0three}) into (\ref{ninefive}) gives
\beqa S(X)&=&\frac{\alpha^2}{g^2}{\rm Tr}\biggl\{-\frac{1}4 F_{\mu\nu}F^{\mu\nu} -\frac 43 i \alpha  (F^{01}  A^2+F^{20}  A^1)+ \frac {2 i\alpha^2}{\sqrt{-\beta}} \Bigl(  \frac 23 -\frac\beta{2\alpha^2}\Bigr)F^{12}A^0   \cr &&\cr &&\qquad + \frac{8 i\alpha}{3}\,([ J_1,A^2]-[ J_2,A^1])A^0- 2  i \alpha\Bigl(  \frac 23 -\frac\beta{2\alpha^2}\Bigr) [A^1,A^2] J_3  \cr &&\cr &&\qquad+\Bigl(\frac \beta 2-\frac{2\alpha^2}3\Bigr)\Bigl((  A^1 )^2+( A^2)^2\Bigr)-{2\alpha^2}\Bigl(\frac \beta{4\alpha^2} -\frac 53\Bigr)( A^0)^2\biggr\}+ S(X|_{\tiny {\rm solution}})\cr &&\label{oneoh5}\eeqa

As stated previously, the commutative limit is obtained by taking $N\rightarrow \infty$, along with $\alpha,\beta\rightarrow 0$ and both $\alpha N$ and $\sqrt{-\beta }\,N$  are finite in the limit.  Upon using  (\ref{eighttwo}) and (\ref{s3ncprtztn}), the commutative limit of the field strengths (\ref{mfsfzysph}) is
\beqa  
 F^{01}&\rightarrow & 2 i\alpha \Bigl(\{j_3, A^1\}-\{j_1, A^0\}- A^2\Bigr)\cr &&\cr
 F^{02}&\rightarrow & -2 i\alpha \Bigl(\{j_3, A^2\}-\{j_2, A^0\}+ A^1\Bigr)\cr &&\cr
 F^{12}&\rightarrow & -2 i\alpha a_0\Bigl(\{j_1, A^2\}-\{j_2, A^1\}- A^0\Bigr)\label{mfsfzysphcl}
 \;,\eeqa where $A^\mu$ are now functions on the commutative sphere.  The trace on functions of the fuzzy sphere is replaced by the corresponding integration on the sphere in the commutative limit.  The relevant integration measure $d\mu(\theta,\phi)$ should be such that the standard  trace identities survive in the limit, i.e., for any three functions $G,H$ and $K$ on the sphere we want $ \int d\mu(\theta,\phi) \{G,H\}K = \int d\mu(\theta,\phi) G\{H,K\}$.  From (\ref{sixty8}) we need to choose the standard integration measure on the sphere $d\mu(\theta,\phi) =\sin\theta d\theta d\phi$ (rather than say $\sqrt{-{\tt g}}\,d\theta d\phi $, where ${\tt g}$ is the determinant of the induced metric). Then  the action (\ref{oneoh5}) reduces  to
\beqa S(X)- S(X|_{\tiny {\rm solution}})&\rightarrow &\frac{2\alpha^4}{g_c^2}\int\sin\theta d\theta d\phi \,\biggl\{-\Bigl(\{j_3, A^1\}-\{j_1, A^0\}\Bigr)^2  -\Bigl(\{j_3, A^2\}-\{j_2, A^0\}\Bigr)^2 \cr &&\cr &&\qquad + a_0^2\Bigl(\{j_1, A^2\}-\{j_2, A^1\}\Bigr)^2 + 2(a_0^2+1)\{j_3,A^1\} A^2
   \cr &&\cr &&\qquad +( A^0)^2-a_0^2\Bigl((  A^1 )^2+( A^2)^2\Bigr)\biggr\}\;,\label{oneoheight}\eeqa where $g_c$ is the commutative limit of the constant $g$.
Following \cite{Iso:2001mg} we write the perturbations $A^\mu$ in terms of gauge potentials $({\cal A}_\theta,{\cal A}_\phi)$ and a scalar field $\psi$ on the sphere using \beqa
A^0&=& {\cal A}_\phi + j_3 \psi  \cr &&\cr
A^1&=& -\sin\phi {\cal A}_\theta -\cot\theta\cos\phi {\cal A}_\phi + j_1 \psi \cr &&\cr
A^2&=& \cos\phi {\cal A}_\theta -\cot\theta\sin\phi {\cal A}_\phi + j_2 \psi\label{Ncpotfrfzy}
\eeqa
Then from the fundamental Poisson bracket (\ref{sixty8}), gauge variations $(\delta{\cal A}_\theta,\delta{\cal A}_\phi)=(\partial_\theta\Lambda,\partial_\phi\Lambda)$ agree with the commutative limit of (\ref{1Ofour}), where  $\Lambda$ is now an infinitesimal function on the commutative sphere.  Substituting (\ref{Ncpotfrfzy}) in (\ref{oneoheight}) gives
\beqa S(X)- S(X|_{\tiny {\rm solution}}) &\rightarrow &\frac{2\alpha^4}{g_c^2}\int\sin\theta d\theta d\phi \,\biggl\{(a_0^2\cot^2\theta-1) {\cal F}_{\theta\phi}^2-\csc^2\theta(\partial_\phi\psi)^2\cr &&\cr  &&\qquad+\Bigl(a_0^2\sin^2\theta-\cos^2\theta\Bigr)(\partial_\theta\psi)^2-\Bigl(3-2(a_0^2+1)\sin^2\theta\Bigr)\psi^2\cr &&\cr  &&\qquad  +2\csc\theta \Bigl((a_0^2+1)\sin^2\theta -2 a_0^2+1\Bigr){\cal F}_{\theta\phi}\psi- 2\cos\theta(a_0^2+1){\cal F}_{\theta\phi}\partial_\theta\psi \biggr\}\;,\cr&&\label{ggsclrft}\eeqa
where ${\cal F}_{\theta\phi}=\partial_\theta{\cal A}_\phi-\partial_\phi{\cal A}_\theta$ is the $U(1)$ gauge field on the surface. 
We remark that the gauge field and scalar field kinetic energies can have opposite signs, a feature that was present in similar two-dimensional systems.\cite{Stern:2014uea},\cite{Stern:2014aqa}  However, gauge fields are nondynamical in two-dimensions. We can solve for ${\cal F}_{\theta\phi}$  from the field equations, yielding
\be {\cal F}_{\theta\phi}=  \frac{\cos{\theta}(a_0^2+1)\partial_\theta\psi-\Bigl((a_0^2+1)\sin^2\theta -2 a_0^2+1\Bigr)\csc\theta\;\psi}{a_0^2\cot^2\theta -1}\;\;+\;\;{\rm constant}\;,\ee
and   substitute back into the action.  Upon setting the constant equal to zero, we get
\beqa S(X)-  S(X|_{\tiny {\rm solution}})&\rightarrow &\frac{2\alpha^4a_0^2}{g_c^2}\int\sin\theta d\theta d\phi \,\biggl\{ \frac{ (\partial_\theta\psi)^2}{(a_0^2+1)\sin^2{\theta}-a_0^2}\,\;-\;\frac{\csc^2\theta}{a_0^2}\,(\partial_\phi\psi)^2\;-\;4m^2_{\rm eff}\psi^2\biggr\}\cr&&\cr &=&\frac{16\alpha^4a_0^2}{g_c^2}\int\sin\theta d\theta d\phi \,\Bigl\{ -\frac 12 \partial^a\psi \partial_a\psi\;-\;\frac 12 m^2_{\rm eff}\psi^2\Bigr\}\;,\label{1twelve}\eeqa
where the index $a=(\theta,\phi)$ is raised and lowered using the induced metric given in (\ref{metricons2}).    The effective mass squared  of the scalar field is $\theta-$dependent
\be m^2_{\rm eff}  =\frac{(a_0^2-1)\Bigl((a_0^2+1)\sin^2{\theta}-3a_0^2\Bigr)}{4a_0^2\Bigl((a_0^2+1)\sin^2{\theta}-a_0^2\Bigr)^2}\label{effmsqs2}\ee
As stated before, the signature of the induced metric is Euclidean when $\sin^2\theta < \frac{a_0^2}{a_0^2+1}$, and  Lorentzian when $\sin^2\theta> \frac{a_0^2}{a_0^2+1}$.   Therefore (\ref{1twelve}) describes a Euclidean field theory for the former and a Lorentzian field theory for the latter.  There are three different possibilities for the Lorentzian field theory: 

\noindent  {\it a)}   The action describes a tachyon when  $a_0^2>1$.  This is since the factor  $(a_0^2+1)\sin^2{\theta}-3a_0^2$  in (\ref{effmsqs2}) is negative in this case.  

\noindent  {\it b)} The scalar field is massless when $a_0^2=1$. 

\noindent  {\it c)} 
The effective mass-squared for the scalar field is positive when\beqa  a_0^2<1 \qquad &{\rm and}& \qquad \frac{ a_0^2}{a_0^2+1} <\sin^2\theta<\frac{3 a_0^2}{a_0^2+1}\eeqa 
It follows that  the action (\ref{1twelve}) describes a massive scalar field throughout the entire Lorentzian region when $\frac 12 \le a_0^2<1$.  On the other hand, when $  a_0^2<\frac 12$ the scalar field becomes tachyonic in the region where $\sin^2\theta >\frac{3 a_0^2}{a_0^2+1}$.

 \section{Concluding Remarks}

We found fuzzy sphere solutions to the Lorentzian IKKT model which provide toy models of a noncommutative two-dimensional closed universe, where time and spatial size  have discrete values.  Singularities in the Ricci tensor appear in the large $N$ (i.e., commutative) limit.  They are analogous to big bang/crunch singularities, with the novel feature that they occur at nonzero spatial size.  Perturbations around the fuzzy sphere solution are described by a scalar field in the commutative limit which can propagate in the Lorentzian region of the manifold.  The scalar field can be massive, massless or tachyonic, the choice depending on the parameter $a_0^2$ (and also on the range of $\theta$ when $a_0^2<\frac 12$).  For  $\frac 12 \le a_0^2<1$ the scalar field is always massive, ensuring the stability of the commutative field theory in this case.  Corrections to the commutative limit are obtained by expressing the matrix product in the action (\ref{oneoh5}) in terms the star product on the sphere\cite{Alexanian:2000uz}-\cite{Balachandran:2005ew}  and keeping the next order terms in the $1/N$ expansion.

For a more realistic model of a noncommutative cosmological space-time, one can look for fuzzy  coset space solutions to the IKKT matrix model associated with dimension $d>2$.\cite{Balachandran:2005ew}  One possible example worth consideration is the fuzzy analogue of the four-dimensional coset $SU(3)/U(2)$.  For  coset spaces with $d>4$ one may be able to make both four-dimensional space-time and extra dimensions noncommutative.  Just as with the example of the fuzzy sphere, the commutative limit may lead to a manifold divided up into regions with different signatures of the metric.  Perturbations about such solutions are expected to be described by a coupled gauge-scalar theory in the commutative limit
A common feature of the emergent field theories in previous examples\cite{Stern:2014uea} is that scalar field and gauge field kinetic energies can appear with opposite sign, which also can be seen in (\ref{ggsclrft}). This sign discrepancy was harmless for $d=2$, since the gauge field could be eliminated.  On the other hand, it is of concern for $d>2$, so it would be interesting to see if this  discrepancy can be cured upon taking the commutative limit of higher dimensional fuzzy coset space solutions.

\bigskip
{\Large {\bf Acknowledgments} }

\noindent
We are very grateful to A.  Pinzul for valuable discussions.

\end{document}